\documentclass[twocolumn,english,conference]{IEEEtran}
\usepackage[T1]{fontenc}
\usepackage{babel}
\usepackage{array}
\usepackage{calc}
\usepackage{amsmath}
\usepackage{amsthm}
\usepackage{amssymb}
\usepackage{graphicx}
\usepackage{cite}
\usepackage{algorithm}
\usepackage{subfigure}
\usepackage{color}

\usepackage{enumitem}
\usepackage{multirow}
\usepackage{hhline}

\usepackage[unicode=true,
bookmarks=true,bookmarksnumbered=true,bookmarksopen=true,bookmarksopenlevel=1,
breaklinks=false,pdfborder={0 0 0},backref=false,colorlinks=false]
{hyperref}
\hypersetup{pdftitle={Your Title},
	pdfauthor={Your Name},
	pdfpagelayout=OneColumn, pdfnewwindow=true, pdfstartview=XYZ, plainpages=false}

\makeatletter

\providecommand{\tabularnewline}{\\}

\let\oldforeign@language\foreign@language
\DeclareRobustCommand{\foreign@language}[1]{%
	\lowercase{\oldforeign@language{#1}}}

\makeatother

\allowdisplaybreaks

\makeatletter
\newcommand{\thickhline}{%
	\noalign {\ifnum 0=`}\fi \hrule height 1.5pt
	\futurelet \reserved@a \@xhline
}
\newcolumntype{"}{@{\hskip\tabcolsep\vrule width 1pt\hskip\tabcolsep}}
\makeatother

\newcolumntype{P}[1]{>{\centering\arraybackslash}p{#1}}
\newcolumntype{M}[1]{>{\centering\arraybackslash}m{#1}}

\begin{document}

	\title{Vulnerability Assessment of Large-scale Power Systems to False Data Injection Attacks}
	
	\author{\IEEEauthorblockN{Zhigang Chu,
			Jiazi Zhang,
			Oliver Kosut, and
			Lalitha Sankar\\}
		\IEEEauthorblockA{School of Electrical, Computer and Energy Engineering\\
			Arizona State University, Tempe, AZ, 85287, USA}}
	\maketitle
	\pagestyle{plain}
	\begin{abstract}
		This paper studies the vulnerability of large-scale power systems to false data injection (FDI) attacks through their physical consequences. An attacker-defender bi-level linear program (ADBLP) can be used to determine the worst-case consequences of FDI attacks aiming to maximize the physical power flow on a target line. This ADBLP can be transformed into a single-level mixed-integer linear program (MILP), but it is numerically intractable for power systems with a large number of buses and branches. In this paper, a modified Benders' decomposition algorithm is proposed to solve the ADBLP on large power systems without converting it to the MILP. Of more general interest, the proposed algorithm can be used to solve any ADBLP. Vulnerability of the IEEE 118-bus system and the Polish system with 2383 buses to FDI attacks is assessed using the proposed algorithm.
	\end{abstract}
	
	\IEEEpeerreviewmaketitle
	\global\long\def\figurename{Fig.}
	\global\long\def\tablename{TABLE}
	
	\section{Introduction}
    Modern electric power systems are cyber-physical systems that work efficiently with integration of real-time monitoring, sensing, communication and data processing. However, this integration makes them vulnerable to cyber-attacks including false data injection (FDI) attacks, wherein a malicious attacker replaces a subset of measurements with counterfeits. FDI attacks can be designed to target system states \cite{Liu2009,Kosut2011}, system topology \cite{Jzhang2016,Liu2017}, and energy markets \cite{Moslemi2018}. Evaluating consequences of FDI attacks often involves solving attacker-defender bi-level linear programs (ADBLPs), wherein the first level models the attacker's objective and limitations (\textit{e.g.}, number of measurements to change), while the second level models the system response under attack via DC optimal power flow (OPF). Examples include attacks that cause line overflows \cite{Liang2015}, locational marginal price (LMP) changes \cite{Jia2014}, operating cost increases \cite{Yuan11} and sequential outages \cite{Che2019}. However, the results have only been demonstrated for small systems. In this paper, we focus on the worst-case FDI attack that causes line overflow as in \cite{Liang2015}, but our goal is to evaluate vulnerability of significantly larger systems (\emph{i.e.} thousands of buses).
	
	The attack design ADBLP can be reformulated as a mixed-integer linear program (MILP) by replacing the second level with its Karush-Kuhn-Tucker (KKT) conditions and rewriting the complementary slackness constraints as mixed-integer constraints. As the system size increases, this MILP becomes harder to solve due to the increasing number of binary variables. In \cite{Chu2016}, we introduced three algorithms, namely row generation (RG) \cite{Dantzig1960}, row and column generation (RCG), and cyber-physical difference maximization (DM), to efficiently solve the ADBLP and evaluate system vulnerability. The first two attempt to reduce the number of binary variables by judiciously eliminating constraints in the second level DCOPF. The third solves a single-level linear program that maximize the difference between the cyber and physical power flows to find bounds of attack consequences.
	
	However, as the system size further scales, RG and RCG will become intractable due to the increasing number of binary variables. For instance, RG becomes intractable on the Polish system with 2383 buses. The DM algorithm may provide loose bounds that are not helpful in understanding system vulnerability. In addition, these three algorithms can only be applied on the FDI attack ADBLP because they are based on the nature of DCOPF and FDI attacks. In this paper, we introduce a modified Benders' decomposition (MBD) \cite{Benders1962} algorithm to solve the ADBLP directly, rather than the re-formulated MILP. The proposed algorithm leverages duality theory to convert the ADBLP into a single level optimization problem, and then applies Benders' decomposition to solve via a sequence of standard linear programs. Since the MBD algorithm does not involve any binary variables, it can be applied on system of any size. Moreover, unlike RG, RCG, and DM algorithms that can only be applied on the attack design ADBLP, the MBD algorithm can be applied on any ADBLP because it is based on general attacker-defender games and is independent of specific details of the second level constraints.
	
	The contributions of this paper are as follows:
	\begin{enumerate}
		\item Introduction of a modified Benders' decomposition algorithm to evaluate vulnerability of large-scale power systems. 
		\item Vulnerability analysis of lines that are susceptible to line overflow attacks. 
		\item Analysis of the impact of overall congestion on vulnerability that helps the system operators better estimate the severity of the attacks.
	\end{enumerate}

	The remainder of this paper is organized as follows. Sec. \ref{sec: SE and models} describes the measurement and attack model. Sec. \ref{sec:OldAttack} summarizes prior work formulating the line overflow attack design ADBLP. Sec. \ref{sec:Methods} introduces the MBD algorithm to solve the ADBLP. Simulation results and concluding remarks are presented in Sec. \ref{sec:Simulation} and \ref{sec:conclusion}, respectively.
	
	\section{\label{sec: SE and models}System Model}
	\subsection{Measurement Model}
	For a power system consists of $n_b$ buses, $n_{br}$ branches, $n_g$ generators, and $n_m$ measurements, the DC measurement model is given by
	\begin{equation}
		z=H_Jx+e\label{eq:DCMeasurement}
	\end{equation}
	where $z$ is the $n_{m}\times1$ measurement vector; $x$ is the $n_b\times1$ vector of bus voltage angles (states); $H_J$ is the $n_{m}\times n_{b}$ measurement Jacobian matrix; $e$ is the $n_{m}\times1$ vector of measurement noise, whose entries are assumed to be jointly distributed as $\mathcal{N}$(0,$R$) where $R=\text{diag}(\sigma_{1}^{2},\sigma_{2}^{2},\ldots,\sigma_{n_{m}}^{2})$.

	\subsection{\label{AttackModel}Attack Model}
	The attacker is assumed to have (i) the ability to perform system-wide DCOPF; and (ii) control of measurements in a subgraph $\mathcal{S}$ of the network.\footnote{While these assumptions may seem unrealistic, we have shown in other work \cite{JZhangGM16} that an attacker can cause comparable physical consequences with much less system knowledge. A similar bi-level optimization problem is introduced to evaluate power system vulnerability to more limited attackers, and the techniques presented in this paper could be readily applied to that problem for large systems. For simplicity, we focus on the stronger attacker in this paper.}

	An $n_{b}\times 1$ attack vector $c \neq 0$ is defined to be \emph{unobservable} if, in the absence of noise, the false measurement $\bar{z}$ created by the attacker satisfies $\bar{z}=z+H_Jc$ \cite{Liu2009}. Let $\hat{x}$ be the estimated states without attack. The residual $r=\bar{z}-H_J(\hat{x}+c)=z+H_Jc-H_J(\hat{x}+c)=z-H_J\hat{x}$ is the same as the residual without the attack. Therefore, this attack can bypass the DC bad data detector (BDD).
	
	For tractability reasons, we use DC power flow model, but the attacks introduced in this paper can also be used to create false data that bypass AC BDD as in \cite{Liang2014}. The system operator will see the results of this unobservable attack as load re-distributions between load buses, while the total load remain unchanged.
	
	\thispagestyle{empty}
	\section{\label{sec:OldAttack}Worst-case Line Overflow Attacks}
	The authors of \cite{Liang2015} introduce an ADBLP that can be re-formulated as an MILP, to determine the worst-case unobservable line overflow attack. The first level models the attacker's objective and limitations, while the second level models the system response via DCOPF. On the IEEE RTS 24-bus system, unobservable attacks found using this MILP are shown to successfully result in generation re-dispatches that cause line overflows.
	
	Instead of modeling the DCOPF problem using the B-$\theta$ method,  this paper models it using the equivalent power transfer distribution factor (PTDF) formulation. With PTDFs, state variables $\theta$ can be eliminated by representing them as a function of the generation dispatches $P_G$, which simplifies the DCOPF to have only $P_G$ as its variable, and can be easier to solve than the B-$\theta$ formulation \cite{Sahraei-Ardakani2017}. Without loss of generality, we assume the flow on the target line is positive; if this is not the case, its absolute value can be maximized.
	
	We formulate the ADBLP as follows:
	\begin{subequations}\label{eq:bilevel}
	\begin{flalign}
	\underset{c}{\text{maximize}}\: \hspace{0.2cm} & P_{l}-\sigma\left\Vert c\right\Vert _{1}\label{eq:Obj1_MaxPF}\\
		\notag \text{subject to}\hspace{0.2cm}\;\\
		& \hspace{-0.9cm}P=\text{PTDF}(G_{B}P_{G}^{*}-P_{D}) \label{eq:Physical_PF}\\
		& \hspace{-0.9cm}\left\Vert c\right\Vert _{1}\le N_{1}\label{eq:con_resources}\\
		& \hspace{-0.9cm}-L_{S} P_{D}\le Hc\le L_{S} P_{D}\label{eq:con_loadshift}\\
		& \hspace{-0.9cm}\left\{P_{G}^{*}\right\} =\text{arg}\left\{ \underset{P_{G}}{\text{min}}\: C_{G}\left(P_{G}\right)\right\} \label{eq:OBJ_MINCOST}\\
		&\notag \hspace{-0.8cm} \text{subject to}\\
		&\hspace{-0.2cm}\begin{array}{lr} 
			\sum_{g=1}^{n_{g}}P_{Gg}=\sum_{i=1}^{n_{b}}P_{Di} & \hspace{0.8cm}(\lambda)\end{array}\label{eq:con_nodebalance}\\
		&\begin{array}{lc}
			\hspace{-0.2cm}-P^\text{max}\le \text{PTDF}(G_{B}P_{G}-P_{D}+Hc)\\\hspace{2.5cm}\le P^\text{max}  \hspace{0.9cm}(F^{\pm})\end{array}\label{eq:con_powerflow}\\
		& \begin{array}{cc}
			\hspace{-0.2cm} P_{G}^{\text{min}}\le P_{G}\le P_{G}^{\text{max}} &\hspace{1.6cm} (\alpha^{\pm})\end{array}\label{eq:con_GENlimit}
	\end{flalign}
	\end{subequations}
	where the variables are:
	\begin{description}[leftmargin=1.8cm,style=multiline]
		\item[$c$] attack vector, $n_{b}\times1$;
		\item[$P$] vector of physical line power flows, $n_{br}\times1$ ;
		\item[$P_l$] physical power flow of target line $l$, scalar;
		\item[$P_{G},P_{G}^{*}$] vectors of generation dispatch variables and optimal
		generation dispatch solved by DCOPF, respectively, both are $n_{g}\times 1$ ;
		\item[$\lambda$] dual variable of the load balance constraint;
		\item[$F^{\pm}, \alpha^{\pm}$] dual variable vectors of line limits and generation limits, respectively;
	\end{description}
	and the parameters are:
	\begin{description}[leftmargin=1.8cm,style=multiline]
		\item[$L_{S}$] load shift factor, in percentage;
		\item[$P_{D}$] vector of real loads, $n_{b}\times1$;
		\item[$N_{1}$] $l_{1}$-norm limit, scalar;
		\item[$H$] dependency matrix between power injection
		measurements and states, $n_{b}\times n_{b}$;
		\item[$G_{B}$] generators to buses connectivity matrix, $n_{b}\times n_{g}$;
		\item[$C_{G}$] generation cost vector, $n_g\times 1$;
		\item[$P^{\max}$] line limits vector, $n_{br}\times1$;
		\item[$P_{G}^{\min},P_{G}^{\max}$]  generation limits vectors, both $n_g\times 1$;
		\item[$\sigma$] penalty of the norm of attack vector $c$, scalar.
	\end{description}
	
	In \eqref{eq:Obj1_MaxPF}, the penalty factor $\sigma$ is a small positive number to limit the attack size; constraint \eqref{eq:Physical_PF} is the physical power flow equation; constraint \eqref{eq:con_resources} models the attacker's limited resources. Ideally, $l_0$-norm should be used to precisely capture the sparsity of $c$, but for tractability reasons we use the $l_1$-norm as a proxy. Constraint \eqref{eq:con_loadshift} limits the percentage of load changes at each bus to avoid detection. DCOPF (2e)--(2h) models the system response to the attack.
	
	The most common approach to solve the ADBLP is to convert it to an MILP (denote \emph{original MILP}) by replacing the second level with its KKT conditions and rewriting the complementary slackness conditions as mixed-integer constraints. This approach suffers from the large number of binary variables when the system size scales. We introduced three algorithms in \cite{Chu2016} to overcome this challenge. The first algorithm uses row generation (RG) to model only a subset of line limits in the second level OPF and yields the optimal attack. The second algorithm utilizes row and column generation (RCG) to judiciously eliminate generation limits in addition to line limits, but it loses optimality guarantee. The third algorithm solves a single-level linear program (LP) maximizing the difference (DM) between target line cyber and physical power flows to provide both lower and upper bounds on attack consequences. Since RG and RCG still involve binary variables, they will become intractable when the system size further scales. The lower and upper bounds provided by DM algorithm may be too loose to assess the severity of the attacks. All these three algorithms are based on the nature of DCOPF and FDI attacks, and hence can only be applied on the FDI attack ADBLP.

	\thispagestyle{empty}
	\section{\label{sec:Methods}The Modified Benders' Decomposition Algorithm}
	In this section, we introduce a modified Benders' decomposition (MBD) algorithm to overcome the shortcomings of the three algorithms introduced in \cite{Chu2016}. This approach is independent of the attack design ADBLP and does not involve binary variables, making it applicable for any ADBLP.  

	Benders' decomposition \cite{Benders1962} can be utilized to solve a linear program with complicating variables in a distributed manner at the cost of iteration \cite{ConejoBook}. It is a popular technique to solve optimization problems of large size or with complicating variables. It is also effective in solving complex optimization problems such as stochastic programs and mixed-integer linear programs. In Benders' decomposition, an optimization problem is decomposed into two sub-problems, wherein variables of each sub-problem are treated as constant in the other. The two sub-problems are solved iteratively until the solution converges. The MBD algorithm is formed by modifying the classic Benders' decomposition algorithm to apply it on any ADBLP without converting it into an MILP.
	
	An ADBLP takes the following form (dual variable of the defender's problem is in parentheses):
	\begin{subequations}\label{general}
		\begin{flalign}
		\underset{u}{\text{minimize}} \hspace{0.17cm} & c_1^Tu+d_1^Tv^* \label{generalLv1Obj}\\
		\notag \text{subject to} & \hspace{0.08cm}\\
		& A_1u \ge b_1 \label{generalLv1Con}\\
		& v^*=\text{arg}\{\underset{v}{\text{min}} \hspace{0.17cm} d_2^Tv\} \label{generalLv2Obj}\\
		\notag &\text{subject to} \hspace{0.12cm} \hspace{0.08cm}\\	  
		& \hspace{1cm}A_2u+A_3v \ge b_2  \hspace{0.97cm} (\beta) \label{generalLv2Con}
		\end{flalign}
	\end{subequations}
	where $u$ and $v$ are the attacker's and defender's decision variables, respectively. The defender has no control on $u$, and hence, $u$ in \eqref{generalLv2Con} is treated as a constant in the defender's problem. The attacker does not directly control $v$, but it controls $v^*$ by changing $u$, assuming it has knowledge of the defender's objective and constraints.
	
	The attack optimization ADBLP \eqref{eq:bilevel} fits in the form of \eqref{general} where the attack vector $c$ is represented by $u$ and DCOPF variable $P_G$, is represented by $v$. In the attacker's objective function, $c_1^Tu$ represents the term $-\sigma \left\Vert c \right\Vert_{1}$, and $d_1^Tv^*$ represents the term $P_{l}$ in \eqref{eq:Obj1_MaxPF}. Equality constraints can be equivalently written as two inequality constraints. For example, \eqref{eq:con_nodebalance} can be written as
	\vspace{0cm}
	\begin{subequations}
		\begin{flalign}
		\textbf{1}^TP_G &\geq \textbf{1}^TP_D\\
		-\textbf{1}^TP_G &\geq -\textbf{1}^TP_D
		\end{flalign}
	\end{subequations}
	which fits the form of \eqref{generalLv2Con}. One can similarly map all the constraints in \eqref{eq:bilevel} to those in \eqref{general}.
	
	The defender's problem \eqref{generalLv2Obj}--\eqref{generalLv2Con}, which represents the system response (DCOPF) to a fixed attack vector, has the following dual problem (note that $u$ is treated as constant here since it is the fixed attack vector from the attacker's problem):
	\vspace{-0.4cm}
	\begin{subequations}
		\begin{flalign}
		\hspace{0.9cm}\underset{\beta}{\text{maximize}} \hspace{0.17cm} & \beta^T(b_2-A_2u)\label{dualObj2ndSP}\\
		\text{subject to} \hspace{0.12cm}& A_3^T\beta=d_2\label{dualCon12ndSP}\\
		& \beta \ge 0.\label{dualCon22ndSP}
		\end{flalign}
	\end{subequations}
	By weak duality \cite{BoydBook}, for any feasible primal/dual pair, the dual objective value is always less than the primal one: 
	\begin{flalign}
	\beta^T(b_2-A_2u) \le d_2^Tv.	\label{dualnature}
	\end{flalign}
	Since the defender's problem is a linear program, it satisfies strong duality. That is, any feasible point $(v,\beta)$ that satisfies
	\begin{flalign}
	\beta^T(b_2-A_2u) \ge d_2^Tv	\label{PlessD}
	\end{flalign}
	is an optimal solution to it. Therefore, constraints \eqref{generalLv2Con}, \eqref{dualCon12ndSP}, \eqref{dualCon22ndSP}, and \eqref{PlessD} guarantee the optimality of the defender's problem, and hence, can be used to convert the ADBLP to a single level problem as:
	\begin{subequations}
		\begin{flalign}
		\hspace{-0.1cm}\underset{u,v,\beta}{\text{minimize}} \hspace{0.17cm} & c_1^Tu+d_1^Tv \label{Bendersobj}\\
		\hspace{-0.1cm} \text{subject to} \hspace{0.08cm} & A_1u \ge b_1 \label{Benderscon1}\\
		& A_2u+A_3v \ge b_2 \label{Benderscon3}\\
		& A_3^T\beta=d_2 \label{Benderscon4}\\
		& \beta^Tb_2-\beta^TA_2u- d_2^Tv \ge 0\label{Benderscon5}\\
		& \beta \ge 0.\label{Benderscon6}
		\end{flalign}
	\end{subequations}
	The bilinear term $\beta^TA_2u$ in \eqref{Benderscon5} is non-convex and thus hard to deal with. To overcome this difficulty, Benders' decomposition is utilized to decompose this optimization problem into two problems, with $u$ as the variable for the main problem (MP) and $v,\beta$ as the variables for the sub problem (SP). 
	The MP takes the following form:
	\begin{subequations}
		\begin{flalign}
		\hspace{-0.25cm}\underset{u,\alpha}{\text{minimize}} \hspace{0.17cm} & c_1^Tu+\alpha \label{MPObj}\\
		\text{subject to} \hspace{0.08cm}
		& A_1u \ge b_1 \label{MPcon1}
		\end{flalign}
	\end{subequations}
	where $\alpha$ is a variable introduced to represent $d_1^Tv$, which will then be updated by adding cuts. 
	The SP is given by:
	\begin{subequations}\label{SP}
		\begin{flalign}
		\underset{v,\beta}{\text{minimize}} \hspace{0.18cm} & d_1^Tv\label{SPObj}\\
		\text{subject to} \hspace{0.11cm}  
		& \beta^Tb_2-d_2^Tv-\beta^TA_2u \ge 0\hspace{0.35cm} (\delta)\\
		& A_3v \ge b_2-A_2u \hspace{1.83cm} (\gamma)\\
		& A_3^T\beta=d_2 \hspace{2.75cm} (\lambda)\\
		& \beta \ge 0.
		\end{flalign}
	\end{subequations}
	At the optimal solution of the SP given by \eqref{SP}, we have
	\begin{flalign}
	d_1^Tv^*=\gamma^Tb_2+\lambda^Td_2-\gamma^TA_2u.\label{SP2Optimal}
	\end{flalign}
	An optimality cut can be added to the MP by taking the right hand side of \eqref{SP2Optimal}:
	\begin{flalign}
	\alpha \ge \gamma^Tb_2+\lambda^Td_2-\gamma^TA_2u. \label{OptCut}
	\end{flalign}
	Note that \eqref{OptCut} is added to the MP, and therefore, $u$ is again a variable. If the SP is infeasible with a given $u$, slack variables $s_i$, $i=1,2,3$, can be introduced to all of the SP constraints to solve the relaxed SP:
	\begin{subequations}\label{SP_relax}
		\begin{flalign}
		\underset{v,\beta, s_i}{\text{minimize}} \hspace{0.18cm} & d_1^Tv\label{SP_relaxObj}\\
		\text{subject to} \hspace{0.11cm}  
		& \beta^Tb_2-d_2^Tv-\beta^TA_2u +s_1\ge 0\hspace{0.35cm} (\hat{\delta})\\
		& A_3v +s_2\ge b_2-A_2u \hspace{1.83cm} (\hat{\gamma})\\
		& A_3^T\beta+s_3=d_2 \hspace{2.75cm} (\hat{\lambda})\\
		& \beta \ge 0.
		\end{flalign}
	\end{subequations}
	where $s_i$, $i=1,2,3$ are the slack variables introduced to ensure feasibility of the relaxed SP. Then, instead of an optimality cut \eqref{OptCut}, a feasibility cut is added to the MP:
	\begin{flalign}
	0 \ge \hat{\gamma}^Tb_2+\hat{\lambda}^Td_2-\hat{\gamma}^TA_2u. \label{FeaCut}
	\end{flalign}
	The MP and SP can then be solved iteratively, with the MP updating $u$ and the SP updating cuts in each iteration.
	\begin{algorithm}[tbh]
		\protect\caption{Modified Benders' Decomposition for Bi-level Linear Programs (MBD)}	\label{alg:Method4}			
		\begin{enumerate}
			\item Set the iteration number $j=1$ and let $u^{(0)}=0$.
			\item Solve the SP \eqref{SP} with $u=u^{(j-1)}$.
			\item If the SP is infeasible, solve the relaxed SP \eqref{SP_relax} and obtain $(\hat{\gamma}^{(j)},\hat{\lambda}^{(j)})$, then add a feasibility cut of form \eqref{FeaCut} to the MP. Otherwise, solve SP \eqref{SP} to get $(v^{(j)},\beta^{(j)},\gamma^{(j)},\lambda^{(j)})$, and add an optimality cut of form \eqref{OptCut} to the MP.
			\item Solve the MP with added cuts and obtain the solution $(u^{(j)},\alpha^{(j)})$.
			\item If $|\frac{d_1^Tv^{(j)}-\alpha^{(j)}}{\alpha^{(j)}}|<\epsilon$, stop. The optimal objective value is obtained as $c_1^Tu^{(j)}+d_1^Tv^{(j)}$. Otherwise, let $j=j+1$ and go to step 2).
		\end{enumerate}
	\end{algorithm}
	
	Solving the SP is equivalent to solving the second level DCOPF under attack \eqref{eq:OBJ_MINCOST}$-$\eqref{eq:con_GENlimit}, while the dual variables of the SP provide information on the objective function \eqref{eq:Obj1_MaxPF}. Since each cut is formulated linearly on the $u$ domain, adding cuts to the MP does not affect its convexity. Thus, MBD is guaranteed to converge in a finite number of iterations \cite{Geoffrion1972}. However, due to the non-convexity of the original ADBLP, global optimal solution cannot be guaranteed \cite{Sahinidis1991}. Therefore, the optimal objective value obtained by MBD, ${P}_{l}^{*(\text{MBD})}$, is a lower bound on $P_{l}^*$, the global optimal objective.
    	
	\thispagestyle{empty}
	\section{\label{sec:Simulation}Simulation results}
	In this section, we present numerical results using the MBD algorithm to solve the FDI attack ADBLP and compare with the RG, RCG, and DM algorithms in \cite{Chu2016}. Table \ref{tab:MethodComparison} lists the key features of all the algorithms, including the original MILP approach. Two test systems are used, namely the IEEE 118-bus system and the Polish system. Before attack, the IEEE 118-bus system and the Polish system have 7 and 17 critical lines, respectively. We exhaustively target all critical lines to assess the vulnerability of these two systems. The $l_1$-norm limit $N_1$ is chosen with increment 0.1 in the range $\left[0.1,1\right]$ for the 118-bus system, and $\left[0.1,2\right]$ for the Polish system. Throughout, Matlab, Matpower, and the Gurobi solver are used to perform the simulations. All tests are conducted using a 3.40 GHz PC with 32 GB RAM.

				\begin{table}[h!]
					\renewcommand{\arraystretch}{1.3}
					\protect\caption{Comparison
						of Four Proposed Algorithms\label{tab:MethodComparison}}
					\centering
					\begin{tabular}{M{3.4cm}|M{0.8cm}|M{1.1cm}|M{1.8cm}}
						\thickhline
						\centering Algorithm & Program Type & Outcome & Tractable Test Cases \tabularnewline
						\hline
						Original MILP & MILP & Optimal Solution & IEEE 24-bus  \tabularnewline
						\hline 
						Row Generation for Line Limit Constraints (RG) & MILP & Optimal Solution & IEEE 24-bus, IEEE 118-bus \tabularnewline
						\hline 
						Row and Column Generation for Line and Generator Limit Constraints (RCG) & MILP & Lower Bound & IEEE 24-bus, IEEE 118-bus, Polish \tabularnewline
						\hline 
						Cyber-physical-difference Maximization (DM) & LP & Lower \& Upper Bounds & IEEE 24-bus, IEEE 118-bus, Polish   \tabularnewline
						\hline
						Modified Benders' Decomposition for Bi-level Programs (MBD) & Iterative LP & Lower Bound & IEEE 24-bus, IEEE 118-bus, Polish  \tabularnewline
						\thickhline 
					\end{tabular}
				\end{table}
	\subsection{\label{Efficiency}Computational Efficiency}
	Table \ref{tab:TimeComparison} illustrates the statistics of the computation time for several target lines using the proposed algorithms with 10\% load shift. For each target line, each algorithm is tested for the full range of $N_1$ values stated above. Note that the number of iterations for MBD varies for different parameter choices (target line, $N_{1\negthinspace}$, and $L_S$), resulting in a large variation in computation time, but overall the efficiency of MBD is comparable to other algorithms.
		\begin{table}[h]
			\renewcommand{\arraystretch}{1.3}
			\protect\caption{Statistics of computation time with 10\% load shift\label{tab:TimeComparison}}
			\centering
			\begin{tabular}{c|c|c|c|c|c}
				\thickhline
				Target line & Algorithm & Max (s) & Min (s) & Avg (s)  & Med (s) \tabularnewline
				\hline
				\multirow{4}{*}{37 of 118-bus} & RG & 7.53 & 0.95 & 3.33 & 1.9\\
                \cline{2-6}
				\multirow{4}{*} & RCG & 1.25 & 0.34 & 0.76 & 0.69\\
				\cline{2-6}
				\multirow{4}{*} & DM & 0.5 & 0.43 & 0.47 & 0.45\\
				\cline{2-6}
				\multirow{4}{*} & MBD & 1.88 & 1.57 & 1.63 & 1.59\\
				\thickhline 
				\multirow{3}{*}{24 of Polish} & RCG & 46.36 & 3.40 & 20.39 & 13.67\\
				\cline{2-6}
				\multirow{3}{*}  & DM & 15.75 & 1.91 & 8.09 & 8.58\\
				\cline{2-6}
				\multirow{3}{*}  & MBD & 12.26 & 10.46 & 11.40 & 11.58\tabularnewline
				\thickhline
				\multirow{3}{*}{292 of Polish} & RCG & 76.34 & 27.47 & 39.29 & 33.69\\
				\cline{2-6}
				\multirow{3}{*} & DM & 16.77 & 1.91 & 7.02 & 6.10\\
				\cline{2-6} 
				\multirow{3}{*} & MBD & 1846.2 & 9.86 & 358.73 & 10.31\tabularnewline
				\thickhline 
			\end{tabular}
		\end{table} 
	
	\subsection{\label{PFresults}Results on Maximal Physical Power Flows} 
\begin{figure}[h]
		\centering{}\includegraphics[trim=0 0.3cm 0 0.3cm, scale=0.76]{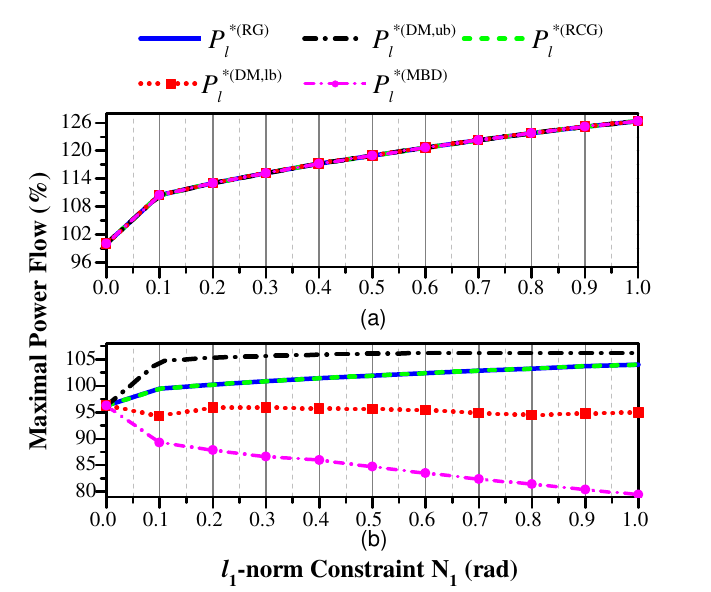}\protect\protect\caption{The maximal power flow vs. the $l_1$-norm constraint ($N_{1}$) with target line (a) 104, and (b) 141 of IEEE 118-bus system. $L_S$=10\%. \label{fig:118_t104}}
\vspace{-0.2cm}
	\end{figure}

	Fig. \ref{fig:118_t104} illustrates the maximal physical power flows with $L_S=10\%$ on target lines 104 and 141 of the IEEE 118-bus system. It demonstrates a comparison of the bounds found by RCG, DM and MBD to the optimal solution provided by RG.
	\vspace{-0.4cm}

	Note that for target line 104 with any $N_1$, all four algorithms yield the optimal solution. For target line 141, we see that $P_l^{*(\text{MBD})}<P_l^{*(\text{DM},\text{lb})}<P_l^{*(\text{RG})}=P_l^{*(\text{RCG})}<P_l^{*(\text{DM},\text{ub})}$, illustrating that $P_l^{*(\text{DM},\text{lb})}$ and $P_l^{*(\text{DM},\text{ub})}$ are not always tight bounds on $P_l^{*}$.
	
\begin{figure}[h]
		\vspace{0.2cm}
		\centering{}\includegraphics[trim=0.55cm 0.3cm 0 0.3cm, scale=0.45]{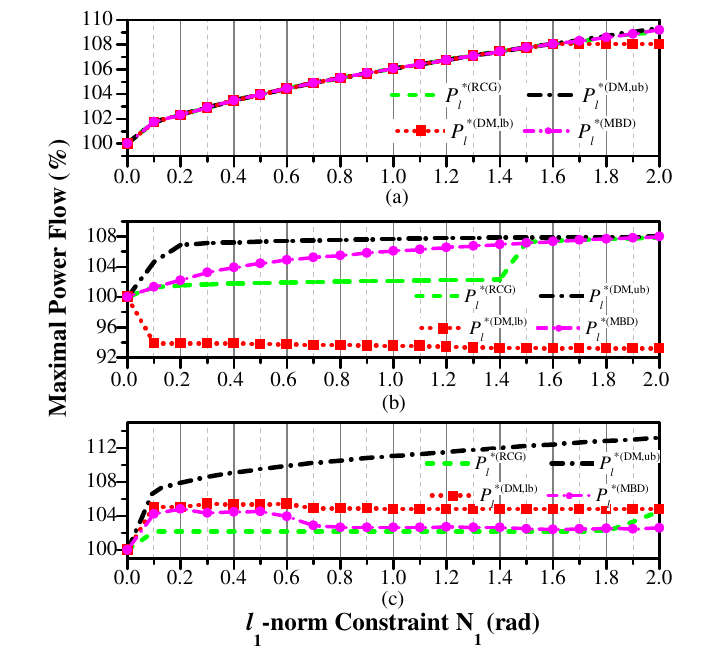}\protect\protect\caption{The maximal power flow vs. the $l_1$-norm constraint ($N_{1}$) with target line (a) 292, (b) 24, and (c) 1816 of the Polish system. $L_S$=10\%. \label{fig:Polish}}
	\vspace{0cm}
	\end{figure}

	The maximal power flows with 10\% load shift for target lines 292, 24, and 1816 of the Polish system are illustrated in Fig. \ref{fig:Polish}. Note that RG is intractable on the Polish system. For target line 292, all three algorithms yield the optimal solution in the range $N_1\in\left[0.1,1.6\right]$, \textit{i.e.,} $P_l^{*(\text{DM},\text{ub})}=P_l^{*(\text{DM},\text{lb})}=P_l^{*(\text{RCG})}=P_l^{*(\text{MBD})}$, but not for the remaining $N_1$. For target line 24, MBD yields the tightest lower bound; while for target line 1816, DM provides the tightest lower bound. 
	
	\subsection{\label{L0result}Results on Attack Resources}
\begin{figure}[h]
		\centering \includegraphics[trim=0 0.3cm 0 0.3cm, scale=0.76]{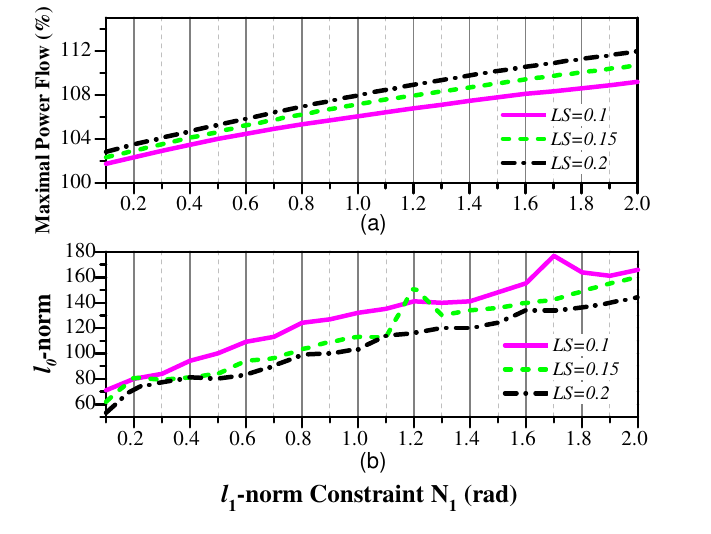}\protect\protect\caption{(a) The maximal power flow and (b) $l_0$-norm of the attack vector vs. the $l_1$-norm constraint ($N_{1}$) for target line 292 of the Polish system. \label{fig:L0LS_t292_Polish}}
		\vspace{-0.3cm}
	\end{figure}

	Fig. \ref{fig:L0LS_t292_Polish} illustrates the relationship between maximal power flow and $l_0$-norm of the attack vector (\emph{i.e.} the number of center buses in the attack) versus the $l_1$-norm constraint $N_1$ for target line 292 of the Polish system, with different load shift constraints. As $N_1$ increases, so does the $l_0$-norm of the attack, indicating that $l_1$-norm is a valid proxy for $l_0$-norm for our problem. If a larger load shift is allowed, the maximal power flow on target line increases, but the resulting $l_0$-norm decreases. This indicates a trade-off between load shift and attacker's resources: as the attacker attempts to avoid detection by minimizing load changes, it will require control over a larger portion of the system to launch a comparable attack. Similar results are also obtained on the IEEE 118-bus system.
	
    \subsection{\label{VulIdentify}Line vulnerability}
\begin{figure}[h]
     		\vspace{0.1cm}
     		\centering \includegraphics[trim=0 0.3cm 0 0.3cm, scale=0.26]{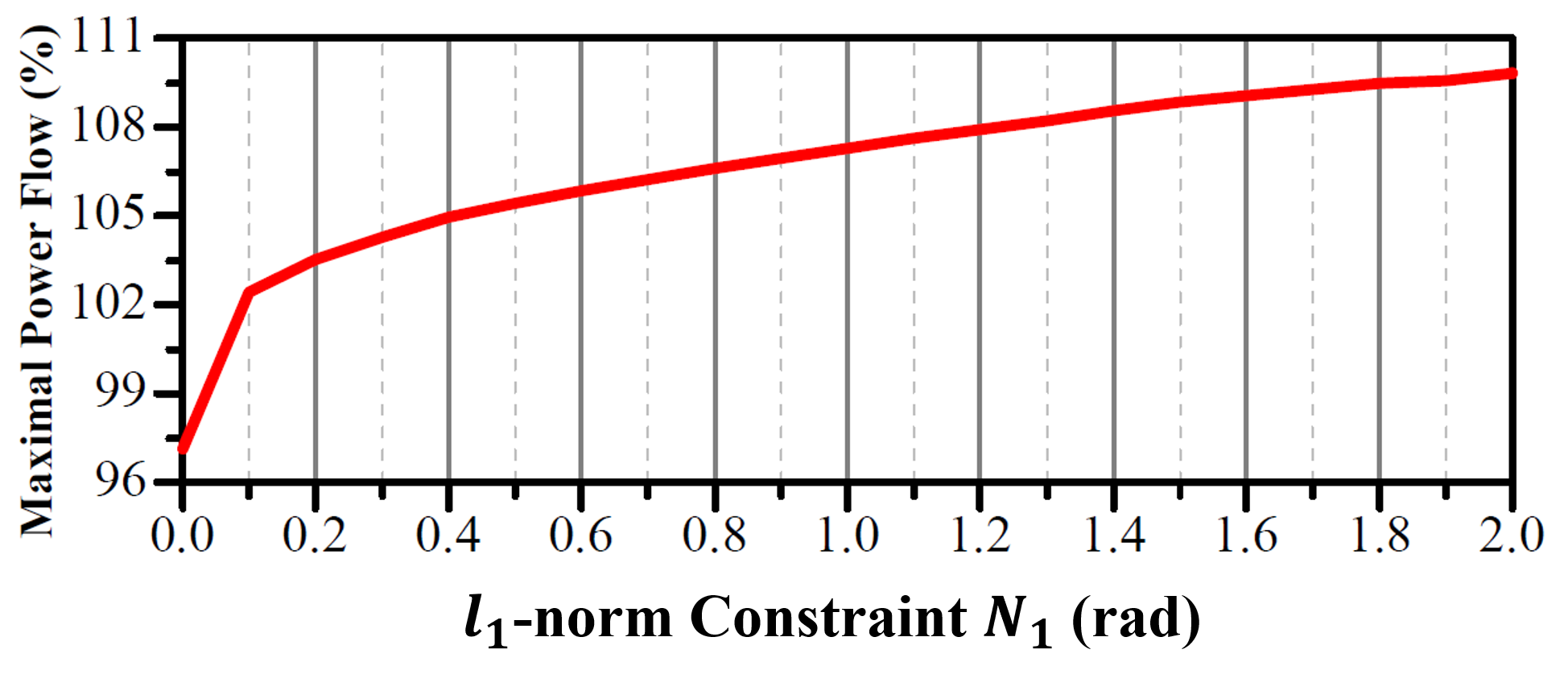}\protect\protect\caption{The maximal power flow vs. the $l_1$-norm constraint ($N_{1}$) for target line 2110 of the Polish system. $L_S$=10\%. \label{fig:t2110_Polish}}
     		\vspace{-0.5cm}
     	\end{figure}

     Since the objective of the attack is to maximize the physical power flow on a target line, it is intuitive that congested lines are more vulnerable to this attack. We have found experimentally that almost every congested line can be overloaded. One exception is line 176 in the IEEE 118-bus system. This is because line 176 is a radial line: it is the only line connected to a bus with a generator and no load. The line limit constraint in the OPF \eqref{eq:con_powerflow} ensures that no possible dispatch could cause the line power flow to exceed the limit, even if based on counterfeit loads. In fact, any line with this radial configuration is immune to the proposed attack; moreover, these radial lines represent the only exceptions to our finding that congested lines can be overloaded. We have also found that lines that are not congested pre-attack may still be vulnerable to this attack, such as line 141 in the IEEE 118-bus system (Fig.\ref{fig:118_t104}(b)) and line 2110 in the Polish system (Fig. \ref{fig:t2110_Polish}).

     \subsection{\label{OverallCongestion}Impact of Overall Congestion on Vulnerability}
     In the above, we have shown that virtually all critical or congested lines are vulnerable to overload. However, the extent of the vulnerability depends on several factors, such as the overall congestion of the system. This phenomenon is illustrated in Fig. 5, which shows the worst-case attack for line 292 of the Polish system under different overall congestion levels. This overall congestion is adjusted by uniformly changing the line ratings for all lines. Note that higher line ratings mean a less congested system. As shown in Fig. 5, as the overall congestion level increases, the maximal power flow on the target line also increases, even though the line is equally congested before attack in each case. 
	\begin{figure}[h!]
		\centering \includegraphics[trim=0cm 0.3cm 1.1cm 0.3cm, scale=0.49]{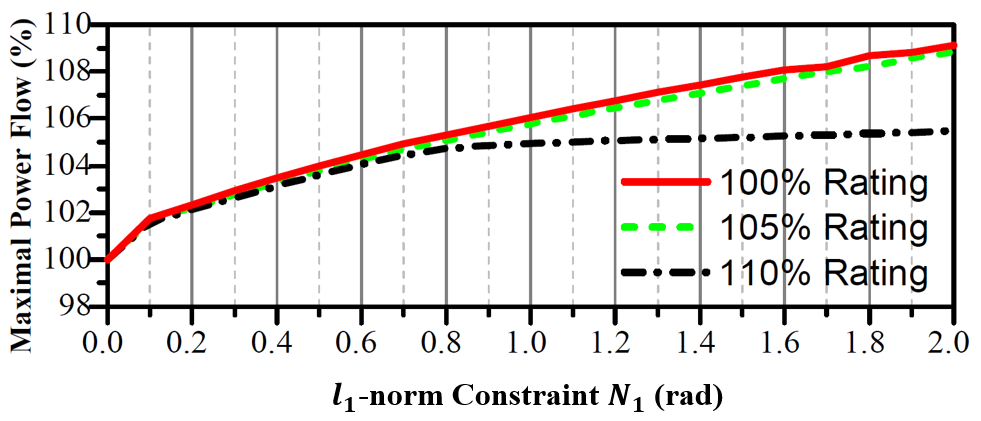}\protect\protect\caption{The maximal power flow vs. the $l_1$-norm constraint ($N_{1}$) for target line 292 of the Polish system under different congestion levels. $L_S$=10\%. \label{fig:overall_con}}
		\vspace{-0.2cm}
	\end{figure}
	\thispagestyle{empty}

	\section{\label{sec:conclusion}Concluding Remarks}
	\vspace{-0cm}
	 We have introduced a modified Benders' decomposition (MBD) algorithm to evaluate the vulnerability of large-scale power systems to FDI attacks. It can be easily applied to any attacker-defender bi-level linear program, making it flexible to evaluate system vulnerability even with additional constraints such as ramp rate constraints, security constraints, and reserve constraints that are common in modern power system operations. Using the MBD algorithm in conjunction with the three algorithms we introduced in \cite{Chu2016} can be helpful in making the system more resilient in the following ways. Using this analysis, the system operators can identify specific lines of vulnerability, and the severity of the attacks. Certain preventive actions can be taken to mitigate attacks. For example, if the system operators find that a line can have overflow under attack, they could artificially reduce the line limit to keep the attack from being successful. Measurements around vulnerable lines can be encrypted to prevent them from being modified. In our optimization problem, the load shift constraint characterizes the detectability of the attack, indicating that load abnormally detectors can help system operators distinguish between natural load changes and possible cyber attacks based on load redistribution.

	\section*{Acknowledgment}
	This material is based upon work supported by the National Science Foundation under Grant No. CNS-1449080 and OAC-1934766.
	\thispagestyle{empty}
	\bibliographystyle{IEEEtran}
	\bibliography{dis}

\begin{thebibliography}{10}
\providecommand{\url}[1]{#1}
\csname url@samestyle\endcsname
\providecommand{\newblock}{\relax}
\providecommand{\bibinfo}[2]{#2}
\providecommand{\BIBentrySTDinterwordspacing}{\spaceskip=0pt\relax}
\providecommand{\BIBentryALTinterwordstretchfactor}{4}
\providecommand{\BIBentryALTinterwordspacing}{\spaceskip=\fontdimen2\font plus
\BIBentryALTinterwordstretchfactor\fontdimen3\font minus
  \fontdimen4\font\relax}
\providecommand{\BIBforeignlanguage}[2]{{%
\expandafter\ifx\csname l@#1\endcsname\relax
\typeout{** WARNING: IEEEtran.bst: No hyphenation pattern has been}%
\typeout{** loaded for the language `#1'. Using the pattern for}%
\typeout{** the default language instead.}%
\else
\language=\csname l@#1\endcsname
\fi
#2}}
\providecommand{\BIBdecl}{\relax}
\BIBdecl

\bibitem{Liu2009}
Y.~Liu, P.~Ning, and M.~K. Reiter, ``False data injection attacks against state
  estimation in electric power grids,'' in \emph{Proceedings of the 16th ACM
  Conference on Computer and Communications Security}, ser. CCS '09, Chicago,
  Illinois, USA, 2009, pp. 21--32.

\bibitem{Kosut2011}
O.~Kosut, L.~Jia, R.~J. Thomas, and L.~Tong, ``{Mali}cious data attacks on the
  smart grid,'' \emph{IEEE Transactions on Smart Grid}, vol.~2, no.~4, pp.
  645--658, 2011.

\bibitem{Jzhang2016}
J.~Zhang and L.~Sankar, ``Physical system consequences of unobservable
  state-and-topology cyber-physical attacks,'' \emph{IEEE Transactions on Smart
  Grid}, vol.~7, no.~4, pp. 2016--2025, July 2016.

\bibitem{Liu2017}
X.~{Liu} and Z.~{Li}, ``Local topology attacks in smart grids,'' \emph{IEEE
  Transactions on Smart Grid}, vol.~8, no.~6, pp. 2617--2626, Nov 2017.

\bibitem{Moslemi2018}
R.~Moslemi, A.~Mesbahi, and J.~M. Velni, ``Design of robust profitable false
  data injection attacks in multi-settlement electricity markets,'' \emph{IET
  Generation, Transmission Distribution}, vol.~12, no.~6, pp. 1263--1270, 2018.

\bibitem{Liang2015}
J.~Liang, L.~Sankar, and O.~Kosut, ``Vulnerability analysis and consequences of
  false data injection attack on power system state estimation,'' \emph{IEEE
  Transactions on Power Systems}, vol.~31, no.~5, pp. 3864--3872, Sept 2016.

\bibitem{Jia2014}
L.~Jia, J.~Kim, R.~J. Thomas, and L.~Tong, ``Impact of data quality on
  real-time locational marginal price,'' \emph{IEEE Trans. Power Systems},
  vol.~29, no.~2, pp. 627--636, 2014.

\bibitem{Yuan11}
Y.~Yuan, Z.~Li, and K.~Ren, ``Modeling load redistribution attacks in power
  systems,'' \emph{Smart Grid, IEEE Transactions on}, vol.~2, no.~2, pp.
  382--390, June 2011.

\bibitem{Che2019}
L.~{Che}, X.~{Liu}, Z.~{Li}, and Y.~{Wen}, ``False data injection attacks
  induced sequential outages in power systems,'' \emph{IEEE Transactions on
  Power Systems}, vol.~34, no.~2, pp. 1513--1523, March 2019.

\bibitem{Chu2016}
Z.~Chu, J.~Zhang, O.~Kosut, and L.~Sankar, ``Evaluating power system
  vulnerability to false data injection attacks via scalable optimization,'' in
  \emph{2016 IEEE International Conference on Smart Grid Communications
  (SmartGridComm)}, 2016, pp. 260--265.

\bibitem{Dantzig1960}
G.~B. Dantzig and P.~Wolfe, ``Decomposition priciple for linear programs,''
  \emph{Operations Research}, vol.~8, pp. 101--111, 1960.

\bibitem{Benders1962}
J.~F. Benders, ``Partitioning procedures for solving mixed-variables
  programming problems,'' \emph{Numerische Mathematik}, no. 4(3), pp. 238--252,
  September 1962.

\bibitem{JZhangGM16}
J.~Zhang, Z.~Chu, L.~Sankar, and O.~Kosut, ``False data injection attacks on
  power system state estimation with limited information,'' in \emph{IEEE PES
  General Meeting}, Boston, MA, July 2016.

\bibitem{Liang2014}
J.~Liang, O.~Kosut, and L.~Sankar, ``Cyber-attacks on {AC} state estimation:
  Unobservability and physical consequences,'' in \emph{IEEE PES General
  Meeting}, Washington, DC, July 2014.

\bibitem{Sahraei-Ardakani2017}
M.~{Sahraei-Ardakani} and K.~W. {Hedman}, ``Computationally efficient
  adjustment of facts set points in {DC} optimal power flow with shift factor
  structure,'' \emph{IEEE Transactions on Power Systems}, vol.~32, no.~3, pp.
  1733--1740, May 2017.

\bibitem{ConejoBook}
A.~J. Conejo, R.~Minguez, E.~Castillo, and R.~Garcia-Bertrand,
  \emph{Decomposition Techniques in Mathematical Programming}.\hskip 1em plus
  0.5em minus 0.4em\relax Springer.

\bibitem{BoydBook}
S.~Boyd and L.~Vandenberghe, \emph{Convex Optimization}.\hskip 1em plus 0.5em
  minus 0.4em\relax Cambridge University Press, 2004.

\bibitem{Geoffrion1972}
A.~M. Geoffrion, ``Generalized {B}enders' decomposition,'' \emph{Optimization
  Theory and Applications}, vol.~10, no.~4, 1972.

\bibitem{Sahinidis1991}
N.~V. Sahinidis and I.~E. Grossmann, ``Convergence properties of generalized
  {B}enders' decomposition,'' \emph{Computers and Chemical Engineering},
  vol.~15, p. 481, 1991.

\end{thebibliography}

\end{document}